# Scientific Understanding and Mathematical Modeling of Emotions of the Spiritually Sublime

*Leonid I. Perlovsky, Harvard University
and Air Force Research Laboratory;* leonid@seas.harvard.edu

*Abstract.* Science strives for a detailed understanding of reality even if this differentiation threatens individual synthesis, or the wholeness of psyche. Religion strives to maintain the wholeness of psyche, even if at the expense of a detailed understanding of the world and Self. This paper analyzes the cognitive forces driving us to achieve both. This analysis leads to understanding emotions of the religiously sublime, which are the foundations of all religions. These seemingly mysterious feelings, which everyone feels, even if rarely, even if without noticing them consciously, even if without being able to name them properly, today can be explained scientifically. And possibly, we may soon be able to measure them in a psychological laboratory. The article briefly reviews new developments in brain imaging that have made new data available, and reviews development and mathematical modeling in cognitive theory explaining these previously mysterious feelings. This new scientific analysis has overcome another long-standing challenge: reductionism. Although religious feelings can be scientifically discussed in terms of concrete neural mechanisms and mathematically modeled, but cannot be reduced to "just this or that" mechanical explanation.

SCIENTIFIC DIFFERENTIATION AND RELIGIOUS SYNTHESIS

Hefner (2008) addressed a fundamental difference between science and religion in terms of strivings for differentiation and synthesis: "…nonspecialists are generalists in their thinking; they work out their own personal, commonsense syntheses that allow for wholeness rather than disjunction of meaning. This cultural pressure for consonance rather than conflict receives very little attention in most religion-science thinking. What are the issues of contrast between this cultural force and the forces for dissonance? Why does our culture embrace both trends?"

Science strives for an increasingly more detailed explanation of the world. Traditional physics explains complex phenomena by "taking them apart" and reducing them to constituent phenomena. This method of analysis has been so successful that it has become identified with the science itself. Complex molecular mechanisms cannot always practically be reduced to atomic interactions. But scientists do not doubt that, given enough effort and computational power, dynamics of complex molecules can be reduced to the interactions among the atoms that make them up. This method of analysis and this way of thinking has been fundamental to scientific and engineering successes from theories of elementary particles and quantum superstrings, to making car engines, airplanes, nuclear bombs, kitchen appliances, and to developing medical devices and pharmaceuticals.

Although there is nothing new in this paragraph the consequences alluded to by Hefner are so deep that I am inclined to repeat it in other words. An engineering design of a kitchen appliance in reality is not always based on explicit detailed *reduction* to interactions among molecules and elementary particles, but this *reduction* is still used when needed in, say, engineering some new materials for a frying pan. More importantly, what is taken for granted here is that whenever engineering "from experience" fails, the analysis can be taken a step deeper, with a complex process taken apart, *reduced* to a more detailed level of reality, and the problem resolved. This is why most large engineering corporations have research divisions, where this *reduction* to deeper levels of scientific analysis is performed.

This process that sometimes is called *differentiation* has become identified with science and with rational thinking. But when applied to religion, differentiation may lead to unsettling conclusions. Religious rituals, when taken apart and logically analyzed, may seem illogical and useless. When differentiation is applied to higher religious experiences and to the highest religious ideas, results of analyses may seem even more bizarre. Some scientists are looking for similarities to experiences of epileptics and schizophrenics or to experiences produced by narcotics. These methods of analyses might serve as steps toward understanding neural substrates involved in religious feelings. But they are also taken by some as indications that religion is "nothing but" a cognitive aberration, etc.

Scientists and commentators looking for arguments that religion is "nothing but" are driven by the human tendency for differentiation, for reducing complex phenomena to simpler ones. These scientists, and to significant extent the whole of science, have ignored the opposite tendency for synthesis, for wholeness. That is why Hefner's comment that human strivings "for consonance rather than conflict receives very little attention" is so important.

Einstein once wrote that he specifically likes thermodynamics, because it was developed without reduction to lower mechanical levels. The world can be understood from basic principles on multiple levels, not just by reduction to the lowest one. Newtonian physics is a classical paradigm of physics, even so it was not built on the theory of superstrings. Physics cannot and should not be built from the lowest to the highest levels of complexity based on the same first principles.

Based on novel developments in cognitive science this paper makes a step toward analyzing a fundamental mechanism responsible for tendency to synthesis. We identify mechanisms for the two fundamental trends, differentiation and synthesis, not only in our culture but in our minds and brains. We will also demonstrate that habits and scientific intuition toward differentiation based on hundreds of years of scientific successes are not applicable to higher levels of the human mind. Differentiation of higher human experiences has fundamental limits and might be scientifically fundamentally flawed, when these limits are ignored.

NEURAL MECHANISMS OF OBJECT PERCEPTION

To perform an analysis of the highest human abilities we begin with a seemingly simple experiment that everyone can perform in three seconds, and which concerns our ability for perception of simple objects, an ability in which we are similar to animals.

Just close your eyes and imagine an object in front of you. Your imagination is vague, not as crisp and clear as with your eyes open. Then open your eyes; the object becomes crisp and clear. It seems to occur momentarily, but actually it takes 1/5th of a second. This is a very long time for neural brain mechanisms – hundreds of thousands of neural interactions. Also note: with opened eyes we are not conscious about initially vague imaginations, we are not conscious about the entire 1/5th of a second, we are conscious only about the end of this process: the crisp, clear object in front of our eyes. Of course, our explanation of this experiment has become simple because through many years of research we have found what goes on in the brain during this 1/5th of a second.

Since the 1950s tens of thousands of scientific publications have been devoted to understanding cognitive mechanisms of the mind and to modeling them mathematically. Not a single one has attempted to explain this simple experiment, and if they would have tried, they would have failed. The main mathematical ideas of algorithms suggested since the 1950s were analyzed in (Perlovsky 1998, 2006): artificial intelligence relying on logical rules; pattern recognition relying on statistical laws; artificial neural networks imitating parallel structures of the brain's neural networks; model systems relying on modifiable models; fuzzy logic attempting to model uncertainties in the world.

After more than fifty years of research and development, computers still cannot do what is easy for young kids or even for animals. Perlovsky (2000, 2006) discusses difficulties faced by each approach and why they would not be able to explain this simple close-open-eyes experiment. These difficulties were manifest as computational complexity (Perlovsky 1998): algorithms that upon logical analyses seemed to be able to solve problems of perception and cognition, practically, even for relatively simple problems, required more computational operations than the number of elementary particle interactions in the entire life of the Universe (this practically infinite number we will call in this paper simply *infinite*). This astronomical computational complexity was related to logic: all of these algorithms relied on logic at some point in their functioning. Even algorithms specifically designed to overcome the limitations of logic, such as neural networks and fuzzy logic, used logic at some point. All learning algorithms use logic in their learning processes. A learning step requires showing an object or its image to a neural network algorithm as follows: "this is a chair"; and this is a logical statement. Fuzzy logic use logic in selecting the degree of fuzziness.

This insurmountable difficulty of logic sits well with the fact that logic is not a fundamental mechanism of cognition. In our close-open-eyes experiment we literally "saw" that crisp (logic-like) perception of an object appears only at the end of the perception process.

In the 1930s mathematical logician Kurt Gödel (1931) proved logically that logic contains irresolvable internal contradictions. The mind resolves its contradictions one way or another; therefore logic could not have been the foundation of cognition. In 2000, New York Times named the five most influential scientists of the past century; along with Einstein, they named Gödel. Gödel's results were recognized as fundamental almost immediately in the 1930s. So, why have generations of mathematicians, cognitive scientists, and philosophers tried to understand cognition as logical operations? The answer is contained in our simple experiment: the mind is not conscious about vague dynamic processes. Consciousness is only aware of logic-like crisp states of mind. Consciousness perceives cognition as a smoothly flowing succession of conscious states, which are separated by hundreds of thousands of neuronal operations. These intertwined dynamics of vague states comprising more than 99% of the mind's operations is almost never accessible to consciousness. This is why even after Gödel's theory received worldwide acclaim as one of the most important results in mathematics, mathematicians, cognitive scientists, and philosophers continued thinking about the mind as a logical system. Our intuitive thinking is built around subjective consciousness.

We have seen that with closed eyes vague states reach consciousness, and although they are not as clear-conscious as with opened eyes, we are still conscious about these imagined objects. In our experiment it was about a simple everyday object at a "low" level of cognition-perception, where a real object was moments away. Sometimes high level cognitive vague states about abstract, life-important ideas reach consciousness, and these experiences might be perceived as miraculous. We discuss these mechanisms later, but let us now return to "simple" perception.

Recently M. Bar, K.S. Kassam, A.S. Ghuman, J. Boshyan, A.M. Schmid, A.M. Dale, M.S. Hamalainen, K. Marinkovic, D.L. Schacter, B.R. Rosen, and E. Halgren (2006) conducted experiments similar to the close-open eyes experiment, but with much more details. It was a culmination of several years of preliminary research. They used functional Magnetic Resonance Imaging (fMRI) to obtain 3-dimensional images of processes in the brain with mm-size resolution. But fMRI is a "slow" technique; it cannot resolve processes in time with the required time resolution (ms). Therefore, fMRI was combined with magneto-encephalography (MEG), measurements of the magnetic field around the head, which cannot produce high-resolution imagery, but provides necessary high temporal resolution of brain activity. Combining these two techniques the experimenters were able to receive high resolution of cognitive processes in space and time.

To explain the novel findings by Bar et al, let me first give a simplified description of the visual perception established through a long line of research. An object image is first projected as in a photo-camera onto the eye retina; from there, through visual nerves the object image is projected onto the visual cortex (in the back of the brain). After processing in the visual cortex it has been thought that an object is recognized in the recognition area (on the side of the brain); this is followed by storage of object information in long-term storage memory (in the front of the brain). In recent decades it has become clear that this picture misses a fundamental aspect of perception: processing of information in the visual cortex involves neural signals from memories about possible objects; signals from memory projected "down" to the cortex are called top-down signals. These images have to be matched to bottom-up images projected from the retina. When the match occurs the mind recognizes the object. This matching process explains some parts of our close-open-eyes experiment: the vague imagination is produced by top-down signals. But details of this process were not known. Bar et al concentrated on three brain areas: visual cortex, object recognition area, and long-term object-information storage area (memory). They demonstrated that memory is activated 130 ms after the visual cortex, but 50 ms *before* the object recognition area. This confirmed that memory of an object is activated before the object is recognized (!) and memory activation produces imagined images. All of these are unconscious. In addition they demonstrated that the imagined image generated by top-down signals from memory to cortex is *vague*, similar to our close-open-eyes experiment. Conscious perception of an object occurs when vague memories become crisp and match a crisp and clear image from the retina, and the object recognition area is activated.

These complex experiment results have a simple interpretation: perception-recognition occurs when the memory of an object matches a concrete object in front of the eyes. But a remembered object, even if similar, is never exactly the same as the one immediately observed: angles, distance, lighting, surrounding objects are always different. Therefore memories have to be vague to have a chance to match anything. Perception requires "improving" knowledge stored in memories, memories have to be fit to concrete conditions in the dynamic process of perception.

Conscious logic-like perceptions occur at the end of a dynamic process "from vague to crisp," from illogical to logical. As discussed, all algorithms considered for modeling of cognition since the 1950s have used logic in one way or another for their operations. They cannot explain why initial "imaginations," top-down signals have to be vague, and they did not model mechanisms, which drive vague images to become crisp, as it occurs when we open our eyes. As a result, they cannot explain cognition even at its simplest "lower" level: perception of everyday objects. One consequence is that computers are not capable of cognition, and another is that cognitive science and philosophy of mind cannot explain cognition. Existing explanations based on logic have faced multiple impasses, as previously mentioned.

The first algorithmic description of perception capable of describing the close-open-eyes experiment as well as the Bar et al experiment was published in (Perlovsky 1987; Perlovsky and McManus 1991), its more contemporary descriptions can be found in (Perlovsky 2000, 2006a). This algorithm is based on a new type of logic, dynamic logic (Perlovsky 2006b, 2007c; Kovalerchuk & Perlovsky 2008, 2009). Whereas usual classical logic describes states (e.g. "this is a chair"), dynamic logic describes processes "from vague to crisp," from vague images, memories, thoughts, decisions, plans to crisp ones. Dynamic logic also overcomes difficulties of artificial intelligence discussed above.

Let me tell a personal story directly related to this discussion. When 20 years ago I came up with the first mathematical algorithms mentioned above, I was afraid to publish these algorithms. As discussed, they were much better than anything previously known for several classical engineering problems. I was sure that thousands of people would immediately pick up the main idea and nobody would remember who the discoverer was. The reality was quite the opposite, the idea has not been "stolen", and it has taken quite a while to be accepted. Now I understand why: it is counterintuitive, because it describes what is unconscious in our minds. These algorithms were ignored for so long by the artificial intelligence and cognitive science communities for the same reason that the Gödelian theory was. Mathematical ideas become quickly popular among mathematicians if they correspond to a sequence of logical steps, this reminds scientists of the conscious operations of mind, the only operations we subjectively know about.

CONCEPTS, INSTINCTS, EMOTIONS

Mind mechanisms corresponding to dynamic logic and to the above discussion have been described in (Perlovsky 2000, 2002, 2006a). Among the main mechanisms are concepts, instincts, and emotions. The mind understands the world in terms of "ideas," similar to the views of Plato and Aristotle. As discussed, these ideas, or concepts, or in contemporary scientific terminology, *models,* are stored in memory. During perception or recognition top-down neural signals project images from models-concepts-memories to the visual cortex. Perception or recognition entails matching these top-down neural signals to bottom-up projections from the eye retina. Since memories never exactly match actual objects, models always have to be modified for perception to occur.

This process of model modification-matching must occur so that we can see the surrounding world. This is a condition for survival, a process that is necessary to satisfy any other instinctual need. Therefore, we have an inborn instinct to fit our memories-models to the world. I call this mechanism the instinct for knowledge. Biologists have been aware of this mechanism since the 1950s (Harlow 1950; Berlyne 1960; Festinger 1957; Cacioppo et al 1996), but only with the help of mathematical modeling, has its fundamental role became clear (Levine and Perlovsky 2008). Since the world is constantly changing, we are constantly refining and revising our models.

The mechanism of the knowledge instinct is similar to other instincts. According to Stephen Grossberg and Daniel Levine (1987), instincts work like internal sensors in the body. These are inborn mechanisms that measure vital body parameters, such as blood pressure, sugar level in the blood, temperature, and dozens of other measures; instincts also indicate to our brain if these parameters are within safe bounds. If they are, we usually do not notice them consciously. If sugar level in the blood goes below a certain point we feel the emotion of hunger.

Several different mechanisms are called emotions; here we refer to the mechanism identified by Grossberg and Levine (1987). Emotions are neural signals that originate in instinctual areas and indicate to cognitive and decision-making areas of the brain satisfaction or dissatisfaction of instinctual needs. For example, if sugar level in the blood is low, emotional neural signals of hunger drive cognitive and decision-making mechanisms to allocate more attention and processing power to finding food, objects that can potentially satisfy instinct for food are recognized faster. These instinctual affects on cognition occur during the dynamic logic process, before recognition occurs; so we are smart not because we understand everything equally well, but because we preferentially understand what we need at every moment.

THE KNOWLEDGE INSTINCT AND AESTHETIC EMOTIONS

The knowledge instinct is similar to other instincts in that our brain has a "sensor" that measures a correspondence or a degree of similarity between models in the mind and objects or situations in the world, and the knowledge instinct drives the mind to maximize this similarity (knowledge). It is different from other 'basic' instincts in that it pertains to processes in the brain, not in the 'lower' body. This paper maintains the scientific point of view that the brain is a part of the body (this is, of course, consistent with monotheism). Still, knowledge and cognition are commonly considered 'higher,' more 'spiritual' functions than eating or sex. In this regard we refer to the knowledge instinct as a 'higher' need than a need for food. Later we argue that it is responsible for all our higher mental abilities.

How do we feel satisfaction or dissatisfaction of the knowledge instinct emotionally? Consider a situation when the surrounding world does not correspond to our knowledge, and surrounding objects behave unexpectedly: doors do not open, teeth cannot bite, a knife spontaneously jumps at you. This is the stuff of thrillers, it is scary, or in a mild dose, disharmonious between knowledge-expectations and reality. Conversely, when objects behave as expected, it is harmonious. Since Immanuel Kant (1790) emotions related to knowledge have been called aesthetic emotions. We would like to emphasize that these emotions, which we feel as harmony-disharmony between our knowledge and the world, are not specific to perception of art, but are inseparable from every act of perception-cognition. Later we demonstrate that aesthetic emotions are related to the beautiful and sublime, but here we call them aesthetic or spiritual for the only reason that they are related to knowledge.

This discussion is central to the topic of this paper, therefore I continue the argumentation. A view of emotions defined by visceral mechanisms (Damasio 1994), as far as discussing higher cognitive functions, seems erroneous in taking secondary effects for the primary mechanisms. People often devote their spare time to increasing their knowledge, even if it is not related to their job and the possibility of promotion. Pragmatic interests could be involved: knowledge makes us more attractive to friends and could help us find sexual partners. Still, there remains the pure joy of finding knowledge, aesthetic emotions satisfying the knowledge instinct. Levine and Perlovsky (2008) discussed brain regions that are likely involved in the knowledge instinct. They also discussed that the knowledge instinct is not the only way people make decisions. Whereas at lower levels of object perception the knowledge instinct acts automatically (otherwise we will not be able to see anything around us), at higher levels of complex and abstract thoughts it is not automatic. Often people don't maximize knowledge; instead they prefer to spare mental efforts and to make decisions by relying on ready-made heuristics or rules. Later we discuss how these two mechanisms interact.

### HIERARCHY OF THE MIND

The mind is organized into an approximate hierarchy (Grossberg 1972) from sensor signals at the bottom, to complex scenes, and abstract ideas higher up. This hierarchy is not strict; it is known that interactions across multiple levels routinely occur during perception and cognition. The perception experiments discussed earlier have demonstrated this. And, although we discussed perception as if concept-models are at the next level to the visual cortex, this is a great simplification. The visual cortex itself is comprised of multiple layers-levels; multiple levels interact across the hierarchy.

Therefore, many neural scientists prefer to call the mind architecture a heterarchy (Grossberg 1972). For simplicity we call it the hierarchy.

It is our hypothesis that understanding of complex abstract concepts involves the same mechanism as object perception: a process of dynamic logic, which fits a vague complex abstract concept-model to specific situations or experiences corresponding to this model. These processes of *differentiation* (of a vague model suitable for many situations into a concrete one for a specific situation) are driven by the knowledge instinct at every level (Perlovsky 2006a). For example, when entering a professor's office, general vague models for a book, shelf, desk, computer, chair, etc. are differentiated into models of concrete objects observed. But this kind of understanding would not take us very far. We strive to understand every situation in its unity, in this case the "professor's office." For this purpose the mind has a corresponding model at a higher level; this higher-level model is remembered from previous encounters with similar situations, and the knowledge instinct drives it to match the current office. This is the process of *synthesis* (of many lower-level models into a higher-level model).

Both *differentiation and synthesis* occur at every level of the hierarchy. I see dozens of different books on my book shelves, and I clearly perceive many of their different features: sizes, colors, titles… The knowledge instinct differentiates my vague "book" model into dozens of crisp and clear models of concrete books. At the same time my knowledge instinct drives me to understand the unity of the "office." This is the process of synthesis: disjointed perceptions of multiple objects are unified in the concept of "office." From the top-down view the mind differentiates vague models into crisp and concrete ones. From the bottom-up view the mind unifies (synthesizes) diverse perceptions into unified models at a higher-level. Differentiation and synthesis are two sides of the knowledge instinct mechanism. Scientific analysis as well as everyday life demands both, understanding of experiences in their details, which is differentiation, and understanding of experiences in their unity, which is synthesis. This analysis is a step toward answering Hefner's question cited at the beginning of this paper: "Why does our culture embrace both trends?" Both trends are most fundamental mind mechanisms, however, the gist of the question is only addressed later, when we consider operations of these mechanisms at the top of the mind hierarchy.

Let us return to the simple close-open-eyes experiment. With closed eyes an imagined object is vague comparative to the same object perceived with opened eyes. But the higher-level abstract concepts cannot be perceived with "opened eyes." At higher levels abstract concept-models cannot be perceived directly by the eyes. This is the reason they are called abstract—concepts like "law," "rationality," or "state," cannot be directly perceived by any senses. Therefore, the entire higher-level cognition has to proceed "with closed eyes," and has to understand the world with vague models. It follows from the previous analysis that higher-level abstract models have to be less conscious than objects in front of our eyes.

This may sound incredible, or at least puzzling: don't we clearly understand concepts like "law" or "state"? To resolve this puzzle we need to understand how cognition interacts with language (Perlovsky 2006a, 2006b, 2007, 2009a). Every concept-model has two parts, linguistic and cognitive, neurally connected to each other. For example, the word "chair" is neurally connected to a cognitive model-image *chair*. Linguistic models are crisp and conscious in the mind, whereas corresponding cognitive models are vague. The higher in the mind hierarchy, the vaguer and less conscious are cognitive models, but language models remain crisp and conscious. At higher levels, language models are like internal "eyes" of the mind. On one hand they provide the *ground* for learning and understanding cognitive models. On the other hand they hide from the mind vague contents of cognitive models. At the level of object perception we can close our eyes and directly experience vague perception models. But at the level of high cognition we cannot easily close the "eyes" of language. Only in the creative process, when inventing new cognitive contents, which have not yet been adequately expressed in language, we may experience vague cognitive models at higher hierarchical levels. Most of the time, we think and perceive complex *cognitive* contents through existing *linguistic* models. Thinking creatively is possible due to the knowledge instinct, which modifies existing vague cognitive models for cognition of new cognitive contents. But this requires a lot of cognitive efforts and the results are vague for a long time until new understanding becomes crisp and is adequately expressed in language. Usually we spare the effort and think in terms of ready-made crisp and conscious linguistic models. Linguistic models may not exactly fit our specific experience, but they are crisp, conscious, and carry the "stamp of approval" of century-old cultural wisdom. Using existing heuristics or rules of thinking and behavior is safe, but it does not advance cultural knowledge. Using the knowledge instinct for creating new cognitive models is risky and uncertain, but this is the process of advancing cultural knowledge (Levine and Perlovsky 2008).

## BEAUTIFUL AND SUBLIME

The mechanism of the knowledge instinct, driving conceptual-emotional understanding, like most mechanisms in our body and mind, is the result of a long evolution. Every one of these mechanisms has evolved for a specific purpose. We are purposeful beings. The purpose of 'basic instincts' is direct survival; the purpose of the knowledge instinct and conceptual understanding of the world is removed a bit from direct survival. At the level of object perception our knowledge instinct is not much different from similar mechanisms in higher animals. But the higher in the hierarchy, the more different the human mind is from the animal mind. Animals do not have language and do not have the complex mind hierarchy (likely these abilities evolved jointly along with dual language-cognitive models; Perlovsky 2006). At higher levels the purpose of the knowledge instinct and conceptual-emotional understanding are far removed from the aims of direct survival.

As discussed, at every level a concept unifies some significant part of lower-level experiences. This is the *purpose* of each concept-model: unify a part of lower-level knowledge. Higher up in the hierarchy there are more general concepts, which unify a larger part of the knowledge and experience. Of course, this synthesis at higher levels is only possible due to omitting lower-level details. Vagueness and lesser consciousness of higher-level concept-models is the "price" our mind pays for generality and unification. This unification or synthesis is an important gain. This *synthesis* or unity of Self is necessary for survival, it is necessary for concentrating the will on the most important goal, alongside the ability for *differentiating* the surrounding world into manifold detailed knowledge.

Concept-models at the top (or near the top) of the hierarchy are mostly vague and unconscious. We cannot consciously perceive their content. Nevertheless we know that synthesis, the unity of "the model at the top" is tremendously important, we know this from clinical cases, when unity of this model is severely destroyed—as in cases of multiple personalities. When psychologists, philosophers, or any one of us discusses *Self* (see McNamara 2009), we talk about some aspects of this concept-model at the top of the mind hierarchy. Is it possible to state definitively what it is? How we feel it? No, because it is not completely within our consciousness, it is outside of what we mean by the conscious "I." Understanding the content of this concept-model, to significant extent, is a cultural construction. One property of this model we know from the above scientific analysis is that the top concept-model has a purpose: to unify our entire experience. Sometimes, rarely, we *feel* this concept-model as a meaning and purpose of our existence.

Is it really possible that our lives have meaning and purpose? I asked this question to several of my friends, who are scientists and do not consider themselves religious. Most often the answer is something like: "Of course not. How can one scientifically discuss such nebulous and vague ideas?" Then I re-phrase my question: "So your life has no more meaning than a rock on the side of the road?" This changes the conversation. No one would agree with this. Most would agree that there is something like a meaning or purpose, but this is so vague, so far away from the possibility of scientific investigation that it is not even clear how to discuss it. I would like to emphasize that this corresponds to the scientific conclusion reached earlier: the top concept-model is vague and mostly unconscious.

Nothing in the surrounding world could directly convince us that life is meaningful. Just the opposite, random deaths and destructions abound, and we know that our material existence is finite. Nevertheless, the feeling of the meaning and purpose in life is so important, that art and religion since time immemorial have been constructing and fortifying this top concept-model in our minds. And sometimes we have a feeling that indeed something like this indeed might be there. Those reading religious and certain philosophical literature might be able to talk at length about the meaning and purpose of life; but as discussed, a language model is only a preparation for a cognitive model, and it might hide from our mind the vagueness and uncertainty of cognitive contents.

Still, sometimes we indeed have an experience that improves our conscious understanding of the contents of the top cognitive model. There is much literature about ecstatic religious experiences and possible neural mechanisms involved (see McNamara 2009). These studies address related problems and could be helpful. However, it is a somewhat separate topic and I would like to exclude it as much as possible from this paper. No one can understand the meaning and purpose of their own life as clearly and consciously as they can understand everyday objects. Yet, rarely we can experience that these top cognitive contents do become clearer, more convincing. Since it is mostly unconscious, conceptual and emotional parts of these contents are not well separated and these experiences are more like feelings than conceptual understandings. These feelings are related to satisfaction of the knowledge instinct at the highest level.

Aesthetic emotions related to satisfaction of the knowledge instinct, as discussed, are experienced as a harmony between our knowledge and the world. At low levels of object recognition they are usually below the level of conscious registration. We do not get elated with harmony because we have correctly perceived, say, a refrigerator, or a cup. We notice these emotions at higher levels, such as when we solve a problem that has preoccupied our mind for a while. The more important and difficult the problem, the more effort it has taken to solve, the stronger the aesthetic emotions related to the achieved new understanding. Conscious efforts to understand what is most important in life could help achieve this understanding, but it is not a finite effort, it may go on indefinitely, and a feeling of some partial understanding from our unconscious may surface unexpectedly. It could occur when thinking about a new physical theory, when going for a walk, when looking through an art catalog or attending a museum. This experience of harmony at the highest level, confirming the meaning and purpose of one's life, we feel emotionally as presence of the beautiful.

This conclusion is not altogether new. Aristotle (IV BCE/1995) has said that the beautiful is unity in manifold, which is an amazingly exact description of the presented theory of the nature of the highest model. The most detailed relations between the beautiful and knowledge possibly have been given by Kant (1790). He called the beautiful "purposiveness without purpose." He perceived this definition as inadequate and tried to improve it in many places in his writings; he explained that purposiveness in the beautiful is not related to any finite aim. But without the notion of the knowledge instinct, without understanding of the dynamical nature of the highest concept-model, which content is culturally constructed and is in the continuous process of refinement and improvement, Kant could not give a positive definition of what is the beautiful. Understanding of the knowledge instinct and that concept-models, including the highest ones, are continuously refined, leads to a conclusion that the beautiful is an aesthetic emotion felt when the knowledge instinct is satisfied at the highest level.

Kantian "purposiveness without purpose" can be refined today by saying that it is a purpose not related to bodily needs, but related to satisfying the instinct for knowledge at the top of the mind hierarchy.

The emotion of sublime, a feeling of spirituality beyond our finite material existence is the foundation of all religions. It is similar to and different from the beautiful. The emotion of the beautiful is related to an improved understanding of the cognitive part of the highest model of life's meaning and purpose. But this understanding is not sufficient for the mind. The mind wants to make this meaning and purpose a part of one's life. In other words, driven by the knowledge instinct, the mind wants to know which behavior would realize this meaning in one's life. Behavior, like understanding, is also governed by models (this is a simplification, but at the high levels in the hierarchy, this simplification is adequate for our purposes). Similar to *cognitive* concept-models, *behavior* concept-models are improved and adapted to concrete circumstances of one's life by the knowledge instinct. When one moves closer to understanding what kind of behavior would realize this highest purpose in his or her life, the emotion of satisfying the knowledge instinct is experienced as spiritually sublime.

## CONTENTS OF THE HIGHEST MODELS

Here we have come close to answering Hefner's question cited at the beginning of the paper. Both differentiation and synthesis are fundamental mechanisms of every person's mind. Life demands use of these mechanisms throughout the lower and middle parts of the hierarchy everyday on multiple occasions. Near the top of the hierarchy, differentiation concerns finding new more detailed and more adequate contents of the highest models, and synthesis concerns finding still higher unifying models. To find the meaning of life everyone has to develop cognitive models in correspondence with cultural models received in language; it is not easy to understand and absorb accumulated cultural wisdom and to develop cognitive models unifying millennial wisdom with today's demands. Even more rare are processes developing new models, which did not yet exist in language and culture. These processes change directions of cultures and they do not occur on a regular basis in the mind of everyone.

The Bible preserved for us many descriptions of these processes; one of the earliest preserved descriptions of a new model of synthesis occurred about 4,000 years ago. At that time writing proliferated, languages were evolving, and consciousness was changing fast (Jaynes 1976). Differentiation overtook synthesis, and stability of cognition and cultures were threatened. Old mechanisms of synthesis did not work, the meaning and purpose of life was disappearing, peoples around the Middle East were losing their bearings, the number of polytheistic gods proliferated, and the region was enmeshed in wars and destruction. God demanded from Abraham that he should leave behind the customs of his people and go out of his land in search of Him. In terms of this paper, Abraham had to find a new synthesis that would be good enough for the increased differentiation of consciousness; he had to find a new model of the highest purpose. According to tradition Abraham found this new synthesis. And today most peoples on the Earth follow in his tradition of a monotheistic religion. Needless to say, for most people feelings of the beautiful and sublime at these highest levels are rare and fleeting, but nevertheless so precious that we usually cherish their memories throughout our lives.

It is worth noting that Jaynes' (1976) analysis of cultural evolution of consciousness and cultures took into account only one of the two fundamental mechanisms of the mind, differentiation, but ignored a need for synthesis. Therefore, his fascinating analysis explained how differentiation was propelled by languages and writings, how it led to loss of assured connection to divinity (ancient synthesis of undifferentiated consciousness), to proliferation of polytheism, to loss of the sense of meaning up to the second millennia BCE, but came to an impasse incapable of explaining the rise of monotheistic religion.

A unified structure of the mind is necessary for survival. Higher animals have significant power of differentiation and are capable of perceiving many diverse objects. Nevertheless they also have a unified executive system in their brains. With evolution of language, acceleration of differentiation and complex mind hierarchy our ancestors had to evolve an instinctual ability for developing synthesis of differentiated consciousness. Otherwise language and concomitant acceleration of differentiation would not be advantageous and would not evolve. These arguments refer to our ancestors living hundreds of thousands or even millions of year ago (Ruhlen 1994). When contemporary language abilities evolved around fifty thousand years ago, instinctual mechanisms for synthesis had certainly been long in place. They continued evolving through genetic and cultural selection and this evolutionary process determined mechanisms for synthesis and unifying abilities of the highest model.

We have significant power over conscious linguistic contents of our highest model, but most of the cognitive contents are unconscious and determined by evolution. Unconscious contents are outside of the conscious "I." Even as the neural brain substrates of this model are within one's brain, a conscious self does not command it, does not "own" it; rather, the opposite relations take place: this model owns and commands one's self at its highest levels. This explains a seeming paradox that a non-religious person, a scientist with materialistic views, would not agree to - the suggestion that principally he is no different than a rock. The unconscious cognitive model at the top of the hierarchy is significantly independent from consciousness and guides consciousness in many ways, in particularly toward feeling its highest purposiveness. This model therefore has the property of an *agent,* independent from one's consciousness, but in control of it. In traditional societies as well as among religious peoples everywhere this is called God.

In our culture, since the ascendance of science, many people consider themselves non-religious. But it is not in one's power to change the unconscious structure of the mind. The model of our highest purposiveness is outside of our conscious control. The scientific analysis in this paper leads to a conclusion that it is not in our power to be "religious" or "irreligious." One could participate in organized religion or refuse to do so. One could consider himself or herself a non-religious person. Or one could choose to study what is known about the contents of the highest models from accumulated wisdom of theologians and philosophers, or by combining this wisdom with scientific methods, as the science-and-religion community does. One can choose to refer to the agency property of the unconscious model at the top of the mind hierarchy, and yet refuse or accept to use the word God.

For a scientist to understand a religious person and vice versa, it is necessary to have common language.

NON-REDUCIBILITY OF SCIENCE

A fundamental difficulty in the past faced by scientists, philosophers, theologians, and religious people when discussing possible scientific explanations of religious experiences and phenomena of consciousness in general has been "reductionism." If a religious experience could be explained scientifically, it seemed the next step would be to reduce this explanation to biology, to chemistry, and to physics. The human being would be no different in principle than a rock, and the same fate would be faced by God. Of course, most people would not tolerate this conclusion. But from a scientific logical viewpoint there was no escape from this conundrum. Some scientists therefore resorted to dualism (Chalmers 1996), refusing to acknowledge that spirit and matter are of the same substance. Most scientists and theologians could not accept this solution either as it contradicts the fundamental premise of monotheism. This conundrum seemed irresolvable.

The reductionism argument was a direct consequence of logic and logic was the foundation of science. There was though a huge hole in this line of reasoning: in the 1930s Gödel (1931/1986) proved that logic is inconsistent, incomplete, and not as logical as expected. But scientists did not know how to use Gödel's results for resolving the problem of reductionism. Roger Penrose (1989, 1994) devoted two books to trying to connect the two and to escape reductionism of consciousness based on Gödel's arguments. Penrose has connected the conundrum of reductionism with another long-standing unresolved problem in basic physics: quantum theory cannot be reconciled with the general theory of relativity. His conclusion was that we would only be able to understand consciousness after new unknown yet basic physical laws of quantum gravity would be discovered. These laws would not be computational (and probably not logical—LP). Many scientists could not accept this as a solution, because it seemed to entail parting with science as we knew it.

The cognitive-mathematical theory described in this paper resolves this conundrum, not by parting with science, but parting with the idea that logic is a fundamental mechanism of the mind. Instead of logic, the fundamental mechanism of the mind is dynamic logic. Dynamic logic, to reiterate, is the process from vague to crisp-logical. Most mind operations are vague, not logical; logical (or almost logical) thoughts, decisions, plans appear at the end of dynamic logic processes. This fact is hidden from our consciousness. Consciousness operates in such a way that we subjectively perceive our mind operations as purely conscious and logical. Our subjective intuition about the mind is based therefore on consciousness and logic. Few scientists were able to overcome this consciousness-logic bias. Among them are Freud, Jung, Zadeh. Theories of unconscious and fuzzy logic that they created are respected, yet too difficult to follow for many, as they require a new type of intuition. The same is true about dynamic logic. Yet, along with unconscious, the dynamic logic mechanism of the mind is confirmed in neuro-imaging experiments, and scientists will have to accept it.

The combination of vague and unconscious mechanisms eliminated the conundrum of reductionism. High level concepts involving the meaning of life are vague and unconscious. We can analyze them and study involved neural mechanisms scientifically. But high level concepts cannot be reduced to finite combinations of constituent simpler concepts. This argument is related to dynamic logic, solving the problem of computational complexity, which was discussed earlier and which is related to Gödel theory, the most fundamental result in mathematics. Computational complexity is due to the fact that high-level decisions involve a choice from an infinite number of combinations of lower-level concepts. Therefore these decisions involve infinite information.

The seemingly unsolvable conundrum of reductionism, which has led many people to doubts about the possibility of combining science and religion, others to dualism, or to postulating future non-computable science, is resolved now. It has become clear that these doubts were based on wrong intuition, on assuming that the mind's main mechanism is logic, that the mind moves in time smoothly from one conscious logical state to another. We know now that conscious logical states of mind are tiny islands among non-logical and unconscious operations, processes of dynamic logic.

FUTURE RESEARCH DIRECTIONS

This paper is one step among many in the science-and-religion effort to mend the schism and unify these two most important endeavors of human spirit. While one fundamental conceptual foundation for the schism seems to have been eliminated by a combination of the new mathematical theory and brain neuro-imaging experimental data, much remains to be done.

A fundamental assumption of this paper is that dynamic logic operates over the entire hierarchy of the mind (Perlovsky 2006). There are no competing theories aimed at explaining the entire hierarchy. Yet, existing neuro-imaging experiments confirm dynamic logic operations only at lower levels of perceptions (Bar et al 2007). Experimental confirmation of dynamic logic operations at higher levels is the next research direction.

The hierarchy of the human mind is possible due to interaction between cognition and language, the neural mechanism of the dual cognitive-language model (Perlovsky 2007). This theory also explains how millennial wisdom accumulated in language is transformed into individual cognition, how collective consciousness is transformed into individual consciousness. Initial experimental evidence confirms interaction between cognition and language (Franklin, Drivonikou, Bevis, Davies, Kay, and Regier 2008). Yet it is but a first step in experimental studies of the dual model operating in the hierarchy.

I referred to the psyche of animals as unified (what I called synthesis of undifferentiated psyche). Is it possible to confirm this experimentally? Are there analogs of multiple personalities among animals? Answers to these questions are not fundamental to the main argument of this paper, but could help to uncover related mechanisms in the human mind.

Neural mechanisms and psychological evidence for spiritually sublime emotions, for religious experiences, and for contents of the highest cognitive models are areas of extensive experimental research (McNamara 2009, 2006; Newberg 2006). The next challenge is connecting this research with the theory in this paper; determine contents of unconscious cognitive models near the top of the hierarchy and correlations of these contents with collective conscious language parts of the models. In particular (McNamara 2009) discusses relations between religious experiences and Self; establishing the role of Self in the highest cognitive models is a next step. A related research direction is to establish the extent to which the top of the hierarchy is governed by the knowledge instinct and to which extent by heuristics and rules (Levine & Perlovsky 2008).

As religion becomes less mystical, science becomes less logical and treads in the areas once reserved for theology.

ACKNOWLEDGEMENTS

It is a pleasure to thank my colleagues discussing with me ideas of this paper, Maxim Frank-Kamenetskii, Mark Karpovsky, Daniel Levine, Lev Levitin, Patrick McNamara, Eugene Shakhnovich, Diana Vinkovetsky, and Yakov Vinkovetsky.

REFERENCES

Aristotle. IV BCE. *Complete Works of Aristotle*, Ed. J. Barnes, Princeton: Princeton Univ. Press, 1995.

Bar, Moshe, K. S. Kassam, A. S. Ghuman, J. Boshyan, A. M. Schmid, A. M. Dale, M. S. Hämäläinen, K. Marinkovic, D. L. Schacter, B. R. Rosen, and E. Halgren. 2006. "Top-down facilitation of visual recognition." *Proceedings of the National Academy of Sciences USA*, 103: 449-54.

Berlyne, Daniel. 1960. *Conflict, Arousal, and Curiosity.* New York, NY: McGraw-Hill.

Cacioppo, John, Richard Petty, J. A. Feinstein, and W. B. G. Jarvis. 1996. "Dispositional Differences in Cognitive Motivation: The Life and Times of Individuals Varying in the Need for Cognition." *Psychological Bulletin* 119:197-253.

Chalmers, David. 1997. *The Conscious Mind: In Search of a Fundamental Theory.* Oxford: Oxford University Press.

Damasio, Antonio. 1994. *Descartes' Error: Emotion, Reason, and the Human Brain.* New York: Grosset/Putnam.

Deming, R., Perlovsky, L.I. (2005). A Mathematical Theory for Learning, and its Application to Time-varying Computed Tomography. New Math. and Natural Computation. 1(1), pp.147-171.

Deming, R.W. and Perlovsky, L.I. (2007). Concurrent multi-target localization, data association, and navigation for a swarm of flying sensors, Information Fusion, 8, pp.316-330.

Deming R., Schindler J., Perlovsky L. (2009). Multitarget/Multisensor Tracking using only Range and Doppler Measurements, IEEE Transactions on Aerospace and Electronic Systems, 45 (2), pp. 593-611.

Festinger, Leon. 1957. *A Theory of Cognitive Dissonance*. Evanston, IL: Row, Peterson.

Franklin, Anna, G. V. Drivonikou, L. Bevis, I. R. L. Davies, Paul Kay, and Terry Regier. 2008. "Categorical perception of color is lateralized to the right hemisphere in infants, but to the left hemisphere in adults." *PNAS*, 10(9): 3221-3225.

Fontanari, J.F. and Perlovsky, L.I. (2004). Solvable null model for the distribution of word frequencies. Physical Review E 70, 042901 (2004).

Fontanari, J.F. and Perlovsky, L.I. (2008). How language can help discrimination in the Neural Modeling Fields framework. Neural Networks, 21(2-3), pp. 250–256

Fontanari, J.F. and Perlovsky, L.I. (2007). Evolving Compositionality in Evolutionary Language Games. IEEE Transactions on Evolutionary Computations, 11(6), pp. 758-769; doi:10.1109/TEVC.2007.892763

Fontanari, J.F. and Perlovsky, L.I. (2008). A game theoretical approach to the evolution of structured communication codes, Theory in Biosciences, 127, pp.205-214.

e-version http://dx.doi.org/10.1007/s12064-008-0024-1.

Fontanari, F.J., Tikhanoff, V., Cangelosi, A., Ilin, R., and Perlovsky, L.I.. (2009). Cross-situational learning of object–word mapping using Neural Modeling Fields. Neural Networks, 22 (5-6), pp.579-585

Gödel, Kurt. 1931/1986. *Kurt Gödel collected works*. Ed. S. Feferman at al. Oxford: Oxford University Press.

Grossberg, Stephen. 1972. "A neural theory of punishment and avoidance, I: Qualitative theory." *Mathematical Biosciences*, 15: 39-67.

Grossberg, Stephen, and Daniel Levine. 1987. "Neural Dynamics of Attentionally Modulated Pavlovian Conditioning: Blocking, Interstimulus Interval, and Secondary Reinforcement." *Applied Optics* 26:5015-5030.

Harlow, Harry. 1953. "Mice, Monkeys, Men, and Motives." *Psychological Review* 60:23-32.

Hefner, Philip. 2008. Editorial. *Zygon: Journal of Religion and Science,* 43(2): 291-296.

Jaynes, Julian. 1976. *The Origin of Consciousness in the Breakdown of the Bicameral mind.* Boston, MA: Houghton Mifflin Co.

Kant, Immanuel. 1790. *Critique of Judgment*, Tr. J. H. Bernard, Macmillan & Co., London, 1914.

Kovalerchuk, B. & Perlovsky, L.I. (2008). Dynamic Logic of Phenomena and Cognition. IJCNN 2008, Hong Kong, pp. 3530-3537.

Kovalerchuk, B. & Perlovsky, L.I. (2009). Dynamic Logic of Phenomena and Cognition. IJCNN 2009, Atlanta; Submitted for journal publication.

Kozma, R., Puljic, M., and Perlovsky, L. (2009). Modeling goal-oriented decision making through cognitive phase transitions, New Mathematics and Natural Computation, 5(1), 143-157.

Levine, Daniel S., Perlovsky, Leonid I. 2008. "Neuroscientific Insights on Biblical Myths: Simplifying Heuristics versus Careful Thinking: Scientific Analysis of Millennial Spiritual Issues". *Zygon*: *Journal of Religion and Science*, 43(4): 797-821.

McNamara, Patrick, Ed. 2006. *Where God and Science Meet: How Brain and Evolutionary Studies Alter Our Understanding of Religion*. Westport, CT: Praeger.

McNamara, Patrick. 2009. *The Neurocognition of religion: A new theory of religion and Self*. Book in preparation.

Newberg, Andrew and Mark Waldman. 2006. Why We Believe What We Believe: Uncovering Our Biological Need for Meaning, Spirituality, and Truth. Boston, MA: Free Press.

Penrose, Roger. 1989. *The Emperor's New Mind*. Oxford: Oxford University Press.

—. 1994. *Shadows of the Mind*. Oxford: Oxford University Press.

Perlovsky, Leonid I. 2000. *Neural Networks and Intellect: Using Model Based Concepts*. New York: Oxford University Press.

—. 1998. "Conundrum of Combinatorial Complexity." *IEEE Trans. PAMI*, **20**(6): 666-670.

—. 2006a. "Toward Physics of the Mind: Concepts, Emotions, Consciousness, and Symbols." *Physics of Life Reviews* 3:23-55.

—. 2006b. "Symbols: Integrated Cognition and Language." In *Semiotics and Intelligent Systems Development*. Eds. R. Gudwin, J. Queiroz. Hershey, PA: Idea Group, pp.121-151.

Perlovsky, L.I. and McManus, M.M. 1991. "Maximum Likelihood Neural Networks for Sensor Fusion and Adaptive Classification." *Neural Networks* 4:89-102.

Perlovsky, L.I. & Marzetta, T.L. (1992). Estimating a Covariance Matrix from Incomplete Independent Realizations of a Random Vector. IEEE Trans. on SP, 40 (8), pp. 2097-2100.

Perlovsky, L.I. (1994). Computational Concepts in Classification: Neural Networks, Statistical Pattern Recognition, and Model Based Vision. Journal of Mathematical Imaging and Vision, 4 (1), pp. 81-110.

Perlovsky, L.I. (1994). A Model Based Neural Network for Transient Signal Processing, Neural Networks, 7(3), pp. 565-572.

Perlovsky, L.I., Coons, R. P., Streit, R.L., Luginbuhl, T.E., Greineder, S. (1994). Application of MLANS to Signal Classification. Journal of Underwater Acoustics, 44 (2), pp.783-809.

Perlovsky, L.I. & Jaskolski, J.V. (1994). Maximum Likelihood Adaptive Neural Controller. Neural Networks, 7 (4), pp. 671-680.

Perlovsky, L.I., Chernick, J.A. & Schoendorf, W.H. (1995). Multi-Sensor ATR and Identification Friend or Foe Using MLANS. Neural Networks, 8 (7/8), pp.1185-1200.

Perlovsky, L.I., Schoendorf, W.H., Tye, D.M., Chang, W. (1995). Concurrent Classification and Tracking Using Maximum Likelihood Adaptive Neural System. Journal of Underwater Acoustics, 45(2), pp.399-414.

Perlovsky, L.I. (1997). Cramer-Rao Bound for Tracking in Clutter and Tracking Multiple Objects. Pattern Recognition Letters, 18(3), pp.283-288.

Perlovsky, L.I. (1997). Physical Concepts of Intellect. Proceedings of Russian Academy of Sciences, 354(3), pp. 320-323.

Perlovsky, L.I., Plum, C.P., Franchi, P.R., Tichovolsky, E.J., Choi, D.S., & Weijers, B. (1997). Einsteinian Neural Network for Spectrum Estimation. Neural Networks, 10(9), pp.1541-46.

Perlovsky, L.I., Schoendorf, W.H., Garvin, L.C., Chang. (1997). Development of Concurrent Classification and Tracking for Active Sonar. Journal of Underwater Acoustics, 47(2), pp.375-388.

Perlovsky, L.I., Schoendorf, W.H., Burdick, B.J. & Tye, D.M. (1997). Model-Based Neural Network for Target Detection in SAR Images. IEEE Trans. on Image Processing, 6(1), pp.203-216.

Perlovsky, L.I. (1998). Conundrum of Combinatorial Complexity. IEEE Trans. PAMI, 20(6) pp. 666-670.

Perlovsky, L.I., Webb, V.H., Bradley, S.R. & Hansen, C.A. (1998). Improved ROTHR Detection and Tracking Using MLANS . AGU Radio Science, 33(4), pp.1034-44.

Perlovsky, L.I. (2000). Beauty and mathematical Intellect. Zvezda, 2000(9), 190-201 (Russian)

Perlovsky, L.I. (2002). Physical Theory of Information Processing in the Mind: concepts and emotions. SEED On Line Journal, 2002; 2(2), pp. 36-54.

Perlovsky, L.I. (2002). Statistical Limitations on Molecular Evolution. Journal of Biomolecular Structure & Dynamics, 19(6), pp.1031-43.

Perlovsky, L.I. (2002). Aesthetics and mathematical theories of intellect. Iskusstvoznanie, 2/02, 558-594 (Russian).

Perlovsky, L.I. (2004). Integrating Language and Cognition. IEEE Connections, Feature Article, 2(2), pp. 8-12.

Perlovsky, L.I. (2006a). Toward Physics of the Mind: Concepts, Emotions, Consciousness, and Symbols. Phys. Life Rev. 3(1), pp.22-55.
Perlovsky, L.I. (2006b). Fuzzy Dynamic Logic. New Math. and Natural Computation, 2(1), pp.43-55.
Perlovsky, L.I. (2006c). Music – The First Priciple. Musical Theatre, http://www.ceo.spb.ru/libretto/kon_lan/ogl.shtml
Tikhanoff. V., Fontanari, J. F., Cangelosi, A. & Perlovsky, L. I. (2006). Language and cognition integration through modeling field theory: category formation for symbol grounding. In Book Series in Computer Science, v. 4131, Heidelberg: Springer.
Perlovsky, L.I. and Deming, R.W. (2007). Neural Networks for Improved Tracking, IEEE Trans. Neural Networks, 18(6), pp. 1854-1857.
Perlovsky, L.I. (2007a). Cognitive high level information fusion, Information Sciences, 177, pp. 2099-2118.
Perlovsky, L.I. (2007b). Evolution of Languages, Consciousness, and Cultures. IEEE Computational Intelligence Magazine, 2(3), pp.25-39
Perlovsky, L.I. (2007c). The Mind vs. Logic: Aristotle and Zadeh. Society for Mathematics of Uncertainty, Critical Review, 1(1), pp. 30-33.
Levine, D.S., Perlovsky, L.I. (2008). Neuroscientific Insights on Biblical Myths: Simplifying Heuristics versus Careful Thinking: Scientific Analysis of Millennial Spiritual Issues. Zygon, Journal of Science and Religion, 43(4), 797-821.
Perlovsky, L.I. (2008). Music and Consciousness, Leonardo, Journal of Arts, Sciences and Technology, 41(4), pp.420-421.
Perlovsky, L.I. (2009a). Language and Cognition. Neural Networks, 22(3), 247-257. doi:10.1016/j.neunet.2009.03.007.
Perlovsky, L.I. (2009b). Language and Emotions: Emotional Sapir-Whorf Hypothesis. Neural Networks, 22(5-6); 518-526. doi:10.1016/j.neunet.2009.06.034
Perlovsky, L.I. (2009c). 'Vague-to-Crisp' Neural Mechanism of Perception. IEEE Trans. Neural Networks, 20(8), 1363-1367.
Perlovsky, L.I. (2010). Musical emotions: Functions, origin, evolution. Physics of Life Reviews, 7(1), 2-27. doi:10.1016/j.plrev.2009.11.001
Perlovsky, L.I. (2010). Neural Mechanisms of the Mind, Aristotle, Zadeh, & fMRI, IEEE Trans. Neural Networks, 21(5), 718-33.
Ilin, R. & Perlovsky, L.I. (2010). Cognitively Inspired Neural Network for Recognition of Situations. International Journal of Natural Computing Research, 1(1), 36-55.
Perlovsky, L.I. & Ilin, R. (2010). Neurally and Mathematically Motivated Architecture for Language and Thought. Special Issue "Brain and Language Architectures: Where We are Now?" The Open Neuroimaging Journal, 4, 70-80. http://www.bentham.org/open/tonij/openaccess2.htm
Perlovsky, L.I. (2010). Intersections of Mathematical, Cognitive, and Aesthetic Theories of Mind, Psychology of Aesthetics, Creativity, and the Arts, 4(1), 11-17. doi: 10.1037/a0018147.
Perlovsky, L.I. (2010). The Mind is not a Kludge, Skeptic, 15(3), 51-55
Perlovsky, L.I. & Ilin, R. (2010). Computational Foundations for Perceptual Symbol System. WCCI 2010, Barcelona. Submitted for journal publication.
Perlovsky, L.I, Bonniot-Cabanac, M.-C., & Cabanac, M. (2010). Curiosity and pleasure, submitted.
Perlovsky, L.I. (2010). Scientific Understanding of Emotions of Religiously Sublime, to be submitted.
Perlovsky, L.I. (2010). Jihadism and Grammars. Comment to "Lost in Translation," Wall Street Journal, June 27, http://online.wsj.com/community/leonid-perlovsky/activity
Ruhlen, Merritt.1994. *The Origin of Language*. New York: John Wiley & Sons, Inc.